\documentclass[
 reprint,
 amsmath,amssymb,
 prl,
]{revtex4-1}

\usepackage{graphicx}
\usepackage{bm}
\usepackage{xcolor}

\begin{document}


\title{Jamming by growth}
\author{Pawel Gniewek}
\thanks{These two authors contributed equally}
\affiliation{Departments of Physics and Integrative Biology, University of California Berkeley, USA 94720}
\author{Carl F. Schreck}
\thanks{These two authors contributed equally}
\affiliation{Departments of Physics and Integrative Biology, University of California Berkeley, USA 94720}
\author{Oskar Hallatschek}
\email{ohallats@berkeley.edu}
\affiliation{Departments of Physics and Integrative Biology, University of California Berkeley, USA 94720}


\begin{abstract}

Growth in confined spaces can drive cellular populations through a jamming transition from a fluid-like state to a solid-like state. Experiments have found that jammed budding yeast populations can build up extreme compressive pressures (over 1MPa), which in turn feed back onto cellular physiology by slowing or even stalling cell growth. Using extensive numerical simulations, we investigate how this feedback impacts the mechanical properties of model jammed cellular populations. We find that feedback directs growth toward poorly-coordinated regions, resulting in an excess number of cell-cell contacts that rigidify cell packings. Cell packings posses anomalously large shear and bulk moduli that depend sensitively on the strength of feedback. These results demonstrate that mechanical feedback on the single-cell level is a simple mechanism by which living systems can tune their population-level mechanical properties.

\end{abstract}

\pacs{Valid PACS appear here}
\maketitle

Granular materials undergo a jamming transition upon compression, at which point the entire system becomes rigid so that further compaction is not possible without pressure build-up~\cite{ohern2002random,ohern2003jamming}. Packings obtained in this manner are spatially-disordered similar to liquids but, like solids, do not yield (irreversibly deform) upon application of an external stress~\cite{liu1998nonlinear}. The transition occurs at a well-defined density $\phi=\phi_J$~\cite{ohern2003jamming}, at which the system is marginally stable ({\it i.e.} removing a single contact causes the system to lose mechanical rigidity)~\cite{wyart2005geometric}. Compression beyond the jamming point ($\phi>\phi_J$) rigidifies packings, resulting in mechanical properties that exhibit nontrivial power law scalings as a function of $\delta\phi=\phi-\phi_J$~\cite{ohern2003jamming,silbert2005vibrations,silbert2006structural,wyart2005geometric,wyart2005rigidity,ellenbroek2006critical}. It has been recently demonstrated that confined microbial populations can similarly drive themselves into a rigid state via cellular growth and division~\cite{delarue2016self}. Cellular populations fundamentally differ from inert granular media in that, whereas granular systems are static unless driven externally~\cite{majmudar2005contact,bi2011jamming,heussinger2009jamming,bohy2012soft,iikawa2016sensitivity,bertrand2016}, cellular populations are active systems that are driven internally as cells consume energy from their environments in order to move or grow~\cite{ramaswamy2010mechanics,bi2015density,delarue2016self, bi2016motility,barton2017model,giavazzi2018flocking,fodora2018statistical}. Growth-driven jamming also differs from the recently studied motility-driven jamming transition~\cite{henkes2011,bi2016motility}, where the system is kept at constant density and is driven by innate cell motility rather than cellular growth. In the case of growth-driven jamming, it is an open question if the cellular packings have the same universal physical properties as conventional granular materials~\cite{ohern2003jamming,wyart2005rigidity}. In particular, experiments have shown that cell-cell forces slow down cell growth~\cite{delarue2016self}, but it remains unknown whether this mechanical feedback at the single-cell level has consequences for population-level mechanical properties. In this work, we show that budding cells can control the mechanical properties of densely-packed populations by leveraging their shape and the coupling between cellular growth rate and cell-cell forces~\cite{delarue2016self,delarue2017scwish}. 

\begin{figure*}
\begin{center}
\includegraphics[width=0.978\textwidth,keepaspectratio]{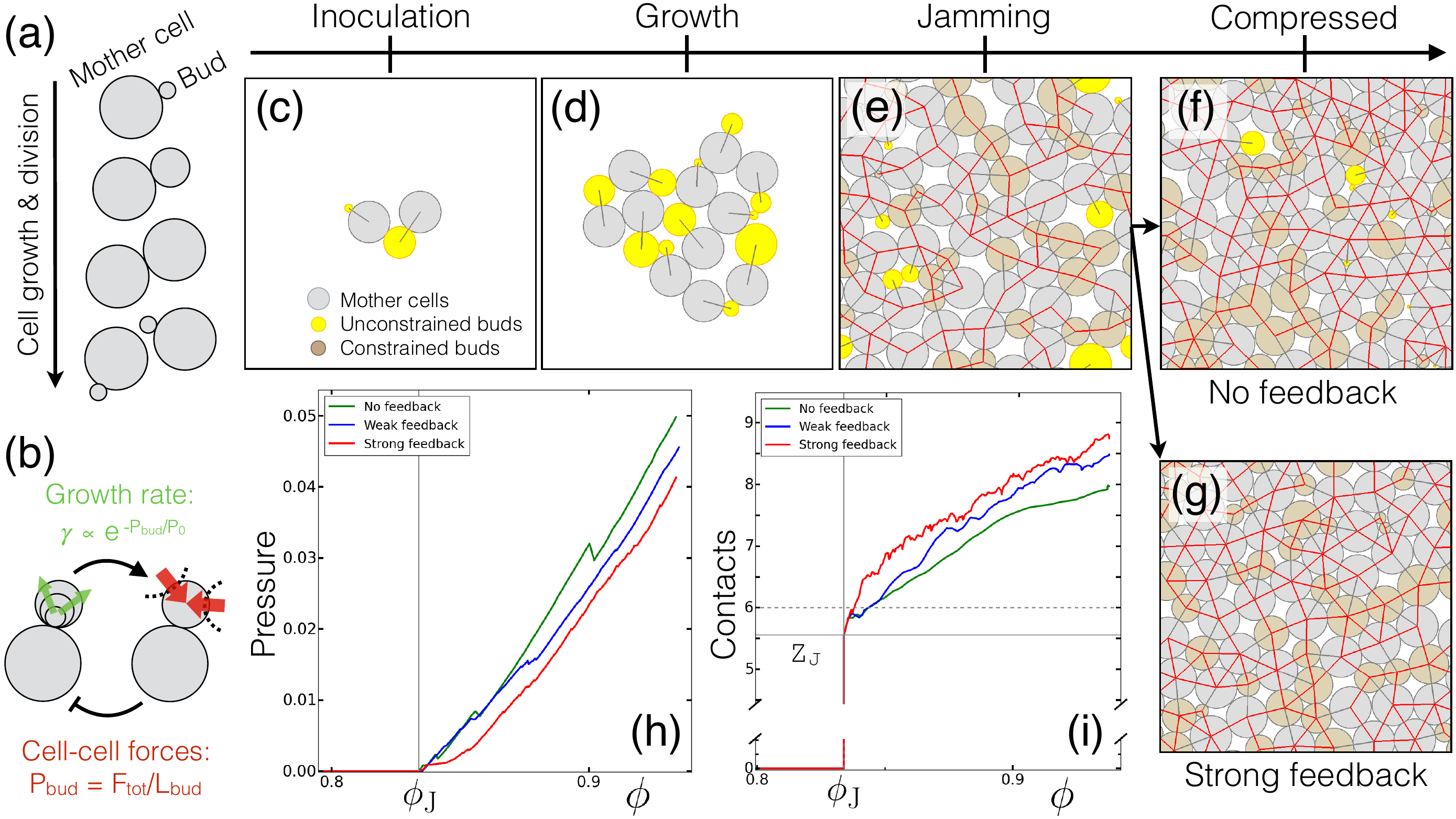}
\caption{(a) Schematic of the growth and division process. Each cell grows by bud expansion. Budding culminates via mother-daughter separation when the daughter reaches the size of a mother cell. (b) Schematic of feedback of cell-cell contact forces onto cell growth. Daughter buds contact their neighbors as they grow (dashed black lines). The associated contact forces generate pressure on the growing bud, defined here as the ratio of total force $F_{\rm tot}$ to the perimeter $L_{\rm bud}$ of the daughter bud (Section~\ref{simmodel}): $P_{\rm bud}=F_{\rm tot}/L_{\rm bud}$. This pressure in turn slows its growth as $\gamma\propto e^{-P_{\rm bud}/P_0}$. (c-g) Snapshots from a typical simulation. (c) Each simulation is inoculated with two cells. (d) Cell growth drives the population to expand outward. During expansion, cells interact with their neighbors via repulsive elastic forces and completely overdamped dynamics. (e) The population undergoes a ``jamming transition'' at $\phi_J=0.84$, at which point a system-spanning network of force-bearing intercellular contacts develops (red lines). At jamming, most mother (gray) and daughter (brown) buds are constrained by their neighbors, but $\approx 25\%$ of buds (yellow) are unconstrained (see SI). Above jamming, populations have fewer unconstrained buds when (f) mechanical pressure feeds back onto cell growth ($P_0=0.001$) than when (g) cellular growth rates are independent of mechanical pressure (both f and g are at $\phi=0.89$). 
(h) The pressure that the entire population exerts on its surroundings $P$ (Section~\ref{P_and_G}) is zero below jamming ($\phi<\phi_J$) and increases as the cells grow above jamming ($\phi>\phi_J$). With no feedback and weak feedback ($P_0=0.005$), $P$ is almost linear in $\phi$. For strong feedback ($P_0=0.001$), $P$ increases more slowly with $\phi$. All pressures are measured in units of the cell-cell modulus $k$ (Section \ref{simmodel}). (i) The number of contacts $Z$ per cell jumps discontinuously from $Z\approx 0$ to $Z=Z_J\approx 5.5$ at jamming at $\phi_J$, and increases more quickly for strong feedback than for weak or no feedback.
(c)-(g) uses box size $L=7\sigma$ and (h), (i) use box size $L=15\sigma$ where $\sigma$ is a cell diameter. (h) and (i) show data for one typical population. 
}
\label{fig:sims}
\end{center}
\end{figure*}

We perform 2D numerical simulations of budding yeast populations growing in space-limited environments. Each cell is represented as conjoined mother and daughter lobes that reproduce asexually via expansion of the daughter ``bud'' (Fig.~\ref{fig:sims}a), a modeling approach first developed in \cite{delarue2016self} alongside microfluidic experiments. In this mode of proliferation, bud expansion progresses until the bud reaches the size of a mother cell, at which point the bud detaches and mother and bud form two new cells. 
To capture the experimentally-measured diminished growth rate under compressive mechanical stress~\cite{delarue2016self}, each cell in our model grows at a rate that decreases exponentially with the pressure exerted on its daughter bud: $\gamma\propto e^{-P_{\rm bud}/P_0}$ (Fig.~\ref{fig:sims}b). The feedback pressure $P_0$ controls the strength of feedback, with smaller values of $P_0$ corresponding to ``stronger'' feedback.

As cells proliferate, repulsive elastic forces between cells (Section \ref{simmodel}, Fig.~\ref{fig:cells}) push the population to expand outward via completely over-damped dynamics (Fig.~\ref{fig:sims}c-g). In the absence of external confinement (Fig.~\ref{fig:sims}c,d), the population remains at zero pressure with no force-bearing contacts between cells. However once the population fills the environment in which it resides, it is driven through a jamming transition (Fig.~\ref{fig:sims}e) at volume fraction $\phi_J\approx 0.84$ that is characterized by a sudden increase in the population pressure $P$ (Fig.~\ref{fig:sims}h) and a discontinuous jump in the number of contacts $Z$  (Fig.~\ref{fig:sims}i). While mechanical feedback does not affect packings below jamming, feedback strength $P_0$ determines how pressure and contacts build up beyond jamming. To understand how mechanical rigidity emerges beyond jamming we first investigate mechanisms underlying the creation of cell-cell contacts, since contacts are know to control the mechanical properties of non-living granular media~\cite{ohern2003jamming,goodrich2012,wyart2005rigidity}.

\begin{figure*}
\begin{center}
\includegraphics[width=0.884\textwidth,keepaspectratio]{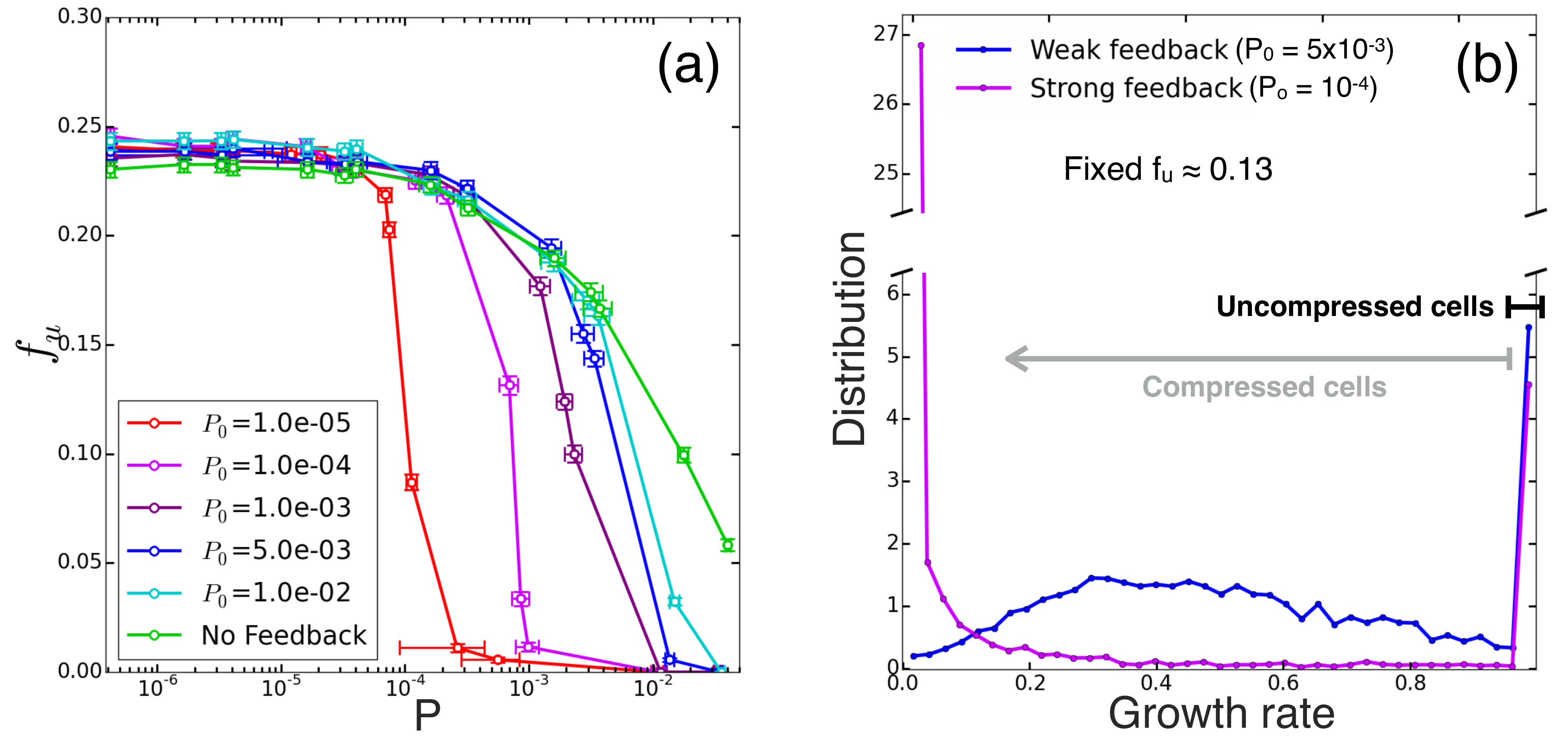}
\caption{(a) Fraction of unconstrained buds $f_u$ as a function of the population pressure generated by growing budding yeast packings above the jamming point. Numerical data is shown for populations with no feedback (bright green line), weak feedback (cyan, blue, purple lines), and strong feedback (magenta and red lines). Populations without feedback have a finite number of unconstrained buds up to $P_{\rm max}\approx 0.1$ which corresponds to $\phi\approx 1$ (Fig.~\ref{fig:sims}h). 
(b) Distribution of cell growth rates for microbial populations with weak ($P_0=5\times 10^{-3}$) and strong ($P_0=10^{-4}$) feedback. In order to measure growth rates as unconstrained buds make contact, both populations have a value of $f_u$ that is $\approx 50\%$ of that measured at jamming ($f_u\approx 0.13$). These values of $f_u$ correspond to $P/P_{\rm max}\approx 0.12$ ($P/P_{\rm max}\approx 0.028$) for weak (strong) feedback. The black bar denotes cells under no or very little pressure, thus growing as they would in the absence of feedback. The gray bar denotes cells whose growth rates are reduced by pressure. The growth rate $\gamma(i)$ of each cell is normalized by $\gamma_i^0$, the growth rate that a cell would have without feedback (Section~\ref{simmodel}). Simulations have box size $L=15\sigma$. Each data point is averaged over $100$ independent inoculations. }
\label{fig:ubuds}
\end{center}
\end{figure*}

At the jamming point, the average number of contacts per cell jumps from $Z=0$ to $Z=Z_J\approx 5.5$ (Fig.~\ref{fig:sims}i), a result that is independent of $P_0$. The value $Z_J\approx 5.5$ is smaller than the naive {\it isostatic} expectation $Z_{\rm iso}^{\rm naive}=6$, predicted by the Maxwell criterion by equating the number of constraints per cell ($Z_{\rm iso}^{\rm naive}/2$) to the number of degrees of freedom per cell ($3$)~\cite{alexander1998}. This deviation from naive isostaticity results from the presence of numerous cells whose buds are not in contact with their neighbors (depicted in yellow in Fig.~\ref{fig:sims}e). Cells with ``unconstrained'' buds are free to rotate about their mother, and therefore correspond to degrees of freedom that are not constrained by cell-cell contacts. By subtracting the number of unconstrained buds per cell $f_u$ from the number cellular degrees of freedom, we can derive a modified isostatic criterion $Z_{\rm iso}=6-2f_u$ (Section \ref{appx:z-iso}) that is satisfied by nearly all simulated populations (Fig.~\ref{fig:Z_iso}). 

We find that a substantial fraction of cells ($f_u\approx 25\%$) have unconstrained buds, which manifests in a strong departure (Fig.~\ref{fig:Z_iso}) from naive isostaticity ($Z_{\rm iso}^{\rm naive}-Z_J\approx 0.5$). The relationship between unconstrained buds and contacts is also observed in non-growing systems. Packings of asymmetric dumbbell-shaped particles that resemble budding cells yield similar results ($f_u\approx 30\%$ and $Z_{\rm iso}^{\rm naive}-Z_J\approx 0.6$)~\cite{VanderWerf2018}, whereas packings of symmetric dumbbells with equal-sized lobes have many fewer unconstrained buds ($f_u\approx 2\%$) and are therefore much closer to isostacity ($Z_{\rm iso}^{\rm naive}-Z_J\approx 0.05$)~\cite{schreck2010comparison}. 

As cells grow beyond the jamming point ($\phi>\phi_J$), the population pressure $P$ builds (Fig.~\ref{fig:sims}h) and unconstrained buds begin to make contact with their neighbors (Fig.~\ref{fig:sims}f,g and Fig.~\ref{fig:ubuds}a). This increase in population pressure, corresponding to comparable pressure on individual cells $\langle P_{\rm bud}\rangle\approx P$, triggers mechanical feedback and slows the growth of cells as $P\gtrsim P_0$ (Fig.~\ref{fig:sims}b). We observe two distinct behaviors for ``strong'' ($P_0/P_{\rm max} \lesssim 0.05$) and ``weak'' ($P_0/P_{\rm max}\gtrsim 0.05$) feedback, where $P_{\rm max}\approx 0.1$ is the pressure felt by populations near confluency $\phi\approx 1$ (see Section~\ref{appx:pmax} for relation of $\phi$ and $P_{\rm max}$). For weak feedback, cell growth rates are not strongly reduced as unconstrained buds make contact with their neighbors (Fig.~\ref{fig:ubuds}b). On the other hand, strong feedback slows the growth of compressed buds by such an extent that it creates two distinct subpopulations: compressed buds that are effectively stalled in their cell cycle and unconstrained (and therefore uncompressed) buds that are actively growing. The threshold between strong and weak feedback corresponds to the pressure ($P/P_{\rm max}\approx 0.05$) at which the majority of previously unconstrained buds contact their neighbors in the absence of feedback (Fig.~\ref{fig:ubuds}a). Therefore, in contrast to weak feedback where cells are driven into contact by nearly uniform population growth, strong feedback directs growth toward unconstrained buds. This directed drives unconstrained buds to make more contacts under strong feedback (Fig.~\ref{fig:sims}f) than in the absence of feedback (Fig.~\ref{fig:sims}g), enhancing the number contacts created per added volume fraction (Fig.~\ref{fig:sims}i). Correspondingly, the preferential growth of unconstrained buds reduces the amount of pressure build-up near jamming (Fig.~\ref{fig:sims}h) because unconstrained buds have free space to grow without incurring cell-cell forces.

\begin{figure*}
\begin{center}
\includegraphics[width=0.898\textwidth,keepaspectratio]{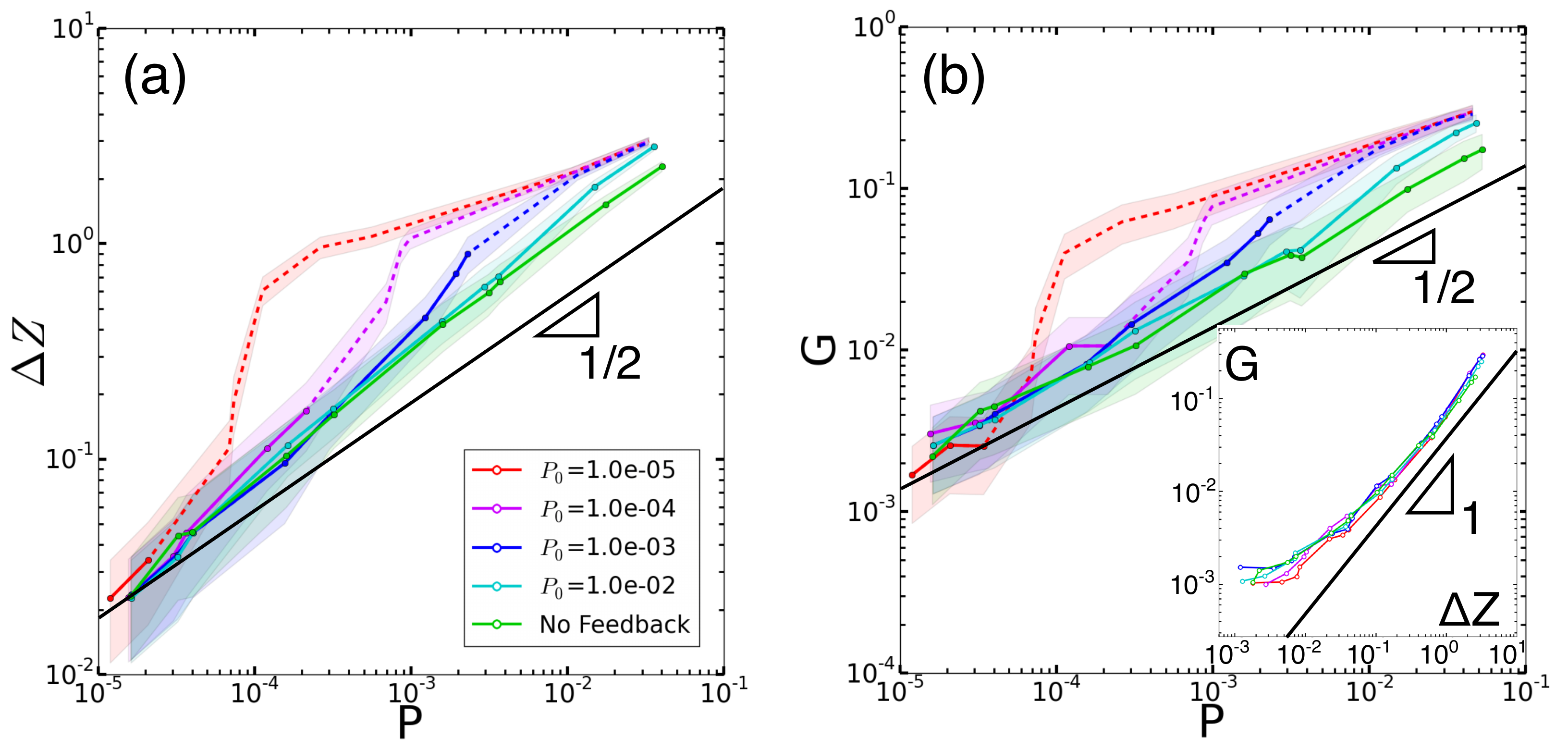}
\caption{(a) Excess number of contacts ($Z$) beyond that measured at jamming ($Z_J$), $\Delta Z=Z-Z_{\rm J}$. Colored lines correspond to growth under a range of feedback values and shaded regions represent one standard deviation. To show where growth has been appreciably slowed by cell-cell forces, dashed colored lines correspond to populations whose average cell growth rate is reduced by a factor of 10 compared to growth without feedback (Fig.~\ref{fig:K_P}). Solid colored lines correspond to growth rates within a factor of 10 of those without feedback. The dotted black line shows the number of contact resulting from unconstrained bud contacts $\Delta Z_u=4f_u\approx 1$ (Section~\ref{appx:z-hyper}). (b) Shear modulus $G$ for cell packings. Inset: Shear modulus in terms of excess contact number $\Delta Z$. Line-types in (b) are the same as shown in (a). Black lines for each panel show known results for disk packings $G\propto\Delta Z\propto P^{1/2}$~\cite{ohern2003jamming}. Simulations have box size $L=15\sigma$. Each data point is averaged over $100$ independent inoculations. 
}
\label{fig:mechanics}
\end{center}
\end{figure*}

By simultaneously driving pressures down and contacts numbers up, strong feedback enables populations to create additional contacts with very little associated pressure build-up compared to growth without feedback (Fig.~\ref{fig:mechanics}a). In the absence of feedback, the excess number of contacts increases roughly as $\Delta Z=Z-Z_J\propto P^{1/2}$, as expected from studies on jamming in non-living systems~\cite{ohern2003jamming,schreck2010comparison}. However, populations growing under strong feedback exhibit abrupt departures from this expectation (Fig.~\ref{fig:mechanics}a) at pressures that vanish for increasing feedback strength ($P\propto P_0$). For strong feedback, additional contacts are generated rapidly as a function of $P$ until all unconstrained buds make contact with their neighbors (Fig.~\ref{fig:ubuds}a), at which point $\Delta Z_{\rm u}\approx 1$ (Fig~\ref{fig:mechanics}b, see Section~\ref{appx:z-hyper} for derivation of $Z_u$). 
This excess of contacts is pushed to lower pressures as $P_0$ decreases, so for $P_0=0$ we expect cell packings to have more contacts than required for mechanical stability even at $P=0$ (i.e., {\it hyperstaticity}). 

How do excess contacts impact the mechanical properties of populations growing under strong feedback? Since prior studies have found that contacts generated by external compression increase the rigidity of granular packings~\cite{ohern2003jamming}, we hypothesize that contacts generated via bud growth likewise rigidify cell packings. To test this hypothesis, we first measure resistance to external compression as quantified by the bulk modulus $B=\phi_e dP/d\phi_e$, where the increase in volume fraction $\phi_e$ is caused compaction rather than cell growth. We find that $B$ increases with feedback strength (Fig.~\ref{fig:B_dPhi_linear}), a direct consequence of the formation of the additional contacts (Fig.~\ref{fig:B_dZ_linear}). In contrast to the increase in $B$ ($dP/d\phi_e$ increases with $P_0$), pressure increases more slowly as volume fraction is added via cellular growth ($dP/d\phi$ decreases with $P_0$ in Fig.~\ref{fig:sims}h). Mechanical feedback therefore allows cell populations to disentangle their mechanical response to internal  perturbations (cell growth) from their response to external perturbations (external compression).

While the generation of excess contacts only slightly modifies the bulk modulus ($B$ increases by $\lesssim 20\%$), we expect these contacts to substantially impact the shear modulus since non-living packings are known to be fragile with respect to shear~\cite{ohern2003jamming,olsson2007,goodrich2012}. By measuring the shear stress $\Sigma_{xy}$ generated under simple shear strain $\gamma$ (Section~\ref{P_and_G}), we find that the shear modulus $G=d\Sigma_{xy}/d\gamma$ scales with pressure as $G\propto P^{1/2}$ in the absence of feedback but increases sharply for strong feedback (Fig~\ref{fig:mechanics}b) as unconstrained buds make contact (Fig~\ref{fig:ubuds}a). The sharp increase in $G$ is indeed controlled by contacts made by unconstrained buds, as we find a one-to-one relationship between contact number and shear modulus (Fig~\ref{fig:mechanics}b inset). The stabilizing role of the added contacts can be understood from constraint counting: both $Z$ and $Z_{\rm iso}$ increase as unconstrained buds make contact, but $Z$ increases faster than $Z_{\rm iso}$ so that packings are pushed above isostaticity as cells grow (Section~\ref{appx:z-hyper}).
The result $G\propto P^{1/2}$ for growth without feedback, also observed for non-living packings~\cite{ohern2003jamming,schreck2010comparison}, suggests that populations near jamming fragile with respect to shear and therefore susceptible to fluidization under thermal excitation~\cite{olsson2007} or cell motility~\cite{henkes2011}. Populations growing under strong feedback, on the other hand, are stabilized by excess contacts even at very small pressure. Therefore, in contrast to populations without feedback and non-living packings where rigidity comes at a cost of increased cell-cell forces, cell populations growing under strong feedback can rigidify themselves with minimal associated pressure. 

We have shown that budding cell populations undergo a growth-driven jamming transition that has mechanical properties not observed in the jamming of non-living systems. Populations growing under mechanical feedback develop a greater number of cell-cell contacts. These contacts are force-bearing and increase the population's resistance to shear and compressive stresses by an amount expected from studies on non-living granular materials~\cite{ohern2003jamming}. As the population grows, this creation of excess intercellular contacts is not accompanied by a faster buildup of the internal pressure in contrast to the anticipated behavior of ordinary granular materials. Thus, the aforementioned feedback mechanism is a simple and efficient mean for expanding microbial populations to increase their resistance to mechanical stress without building up growth-limiting compressive mechanical forces. This mechanism may have important biological consequences for growing microbial populations, such as the increased resistance to mechanical stress may prevent unwanted fluidization that can be caused by processes such as division and apoptosis~\cite{Ranft2010,matozfernandez2017nonlinear,jacobeen2018cellular,jacobeen2018cellular,jacobeen2018geometry} or cell motility~\cite{bi2015density,henkes2011}.

{\it Acknowledgments.} Research support is provided by the Simons Foundation Award No. 327934, NSF Career Award No. 1555330, General Medical Sciences of the NIH Award No. R01GM115851. This work also benefited from the facilities and staff of the National Energy Research Scientific Computing Center, a DOE Office of Science User Facility supported by the Office of Science of the U.S. DOE under Contract No. DE-AC02-05CH11231. We thank J. Paulose and M. Duvernoy for helpful discussions throughout the course of this work.


%

\clearpage

\setcounter{figure}{0}  
\renewcommand{\thefigure}{S\arabic{figure}}    
\section{Supplementary Information}

The source-codes are available on GitHub \cite{sourcecodes}.

\subsection{Cell-based simulations}\label{simmodel}
\begin{figure*}
\includegraphics[width=\textwidth,keepaspectratio]{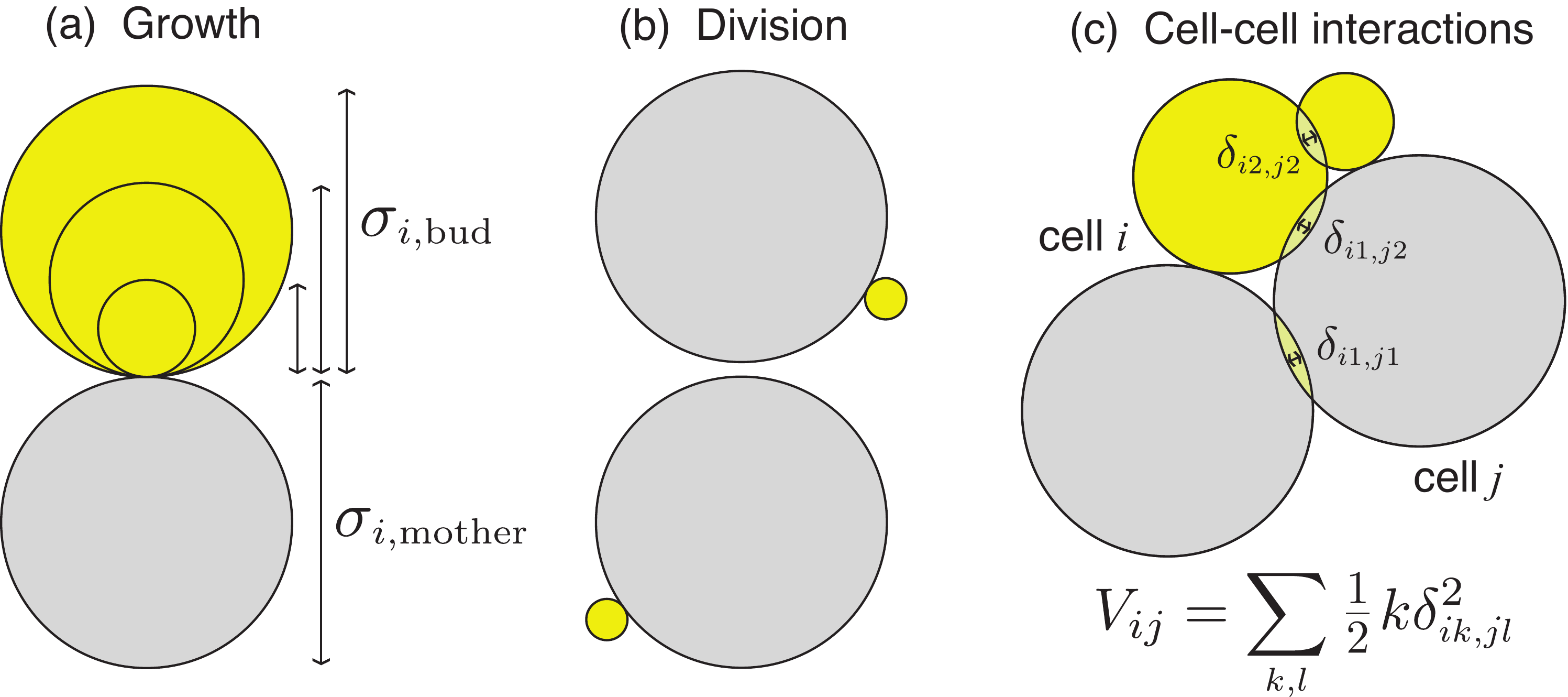}
\caption{
Schematic of (a) the growth and (b) division processes and (c) cell-cell interactions in our cell-based simulations. Each cell is composed to two lobes, the mother (gray) and bud (yellow). (a) During growth, the mother lobe diameter of cell $i$ stays fixed at $\sigma_{i,{\rm mother}}=\sigma$ while the bud grows from $\sigma_{i,{\rm bud}}=0$ to $\sigma_{i,{\rm mother}}=\sigma$. (b) Once the bud reaches $\sigma_{i,{\rm mother}}=\sigma$, cell $i$ divides into two new daughter cells that have random orientations. (c) Cells $i$ and $j$ interact only upon overlap ($\delta_{ik,jl}$) via repulsive linear spring interactions with modulus $\rm k$.}
\label{fig:cells}
\end{figure*}

In 2D cell-based simulations, illustrated in Fig.~\ref{fig:cells}, cells are modeled as two frictionless rigidly-attached spherical lobes~\cite{schreck2010comparison,delarue2016self} (mother and bud) that grow exponentially in time at rate $\gamma_i$ by bud expansion (Eq. 1) and interact via repulsive spring forces with elastic modulus $\rm k$  (Eq. 2):

\begin{equation}\label{eq:growth}
\dot{a}_i = \gamma_i a_i
\end{equation}

\begin{equation}\label{eq:energy}
\rm V = \sum_{i>j}\sum_{k,l}\frac{1}{2} k\delta_{ik,jl}^2\Theta(\delta_{ik,jl})
\end{equation}
where $a_i=\frac{\pi}{4}(\sigma_{i,{\rm mother}}^2+\sigma_{i,{\rm bud}}^2)$ is the cell area, $\sigma_{i,\rm mother}$ ($\sigma_{i,\rm bud}$) is the diameter of the mother (bud), $\rm V$ is the total potential energy,  $\delta_{ik,jl}=\frac{1}{2}\big(\sigma_{ik}+\sigma_{jl}\big)-\big|{\bf r}_{ik}-{\bf r}_{jl}\big|$ is the overlap between lobes $k$ of cell $i$ and $l$ of cell $j$, and $\Theta$ is the Heaviside Step function (Fig.~\ref{fig:cells}). The potential energy is measured in units of $\rm k\sigma_{i,{\rm mother}}^2$ throughout this paper. 

The system is relaxed with quasistatic dynamics via conjugate gradient energy minimization \cite{press1992numerical}. Conjugate gradient minimization terminates upon one of the two conditions: (i) two successive steps j and j+1 yield nearly the same energy value $(\rm V_{j+1} - V_j) / V_j < \epsilon_{tol}^1 = 10^{-16}$ or (ii) the potential energy per particle at the current step is $\rm V_j/(kN) < \epsilon_{tol}^2 = 10^{-16}$. The forces (${\bf F}_i$) and torques ($T_i$) acting on each cell are calculated as:
\begin{equation}
{\bf F}_i=-\nabla_{{\bf r}_i} V
\end{equation}
\begin{equation}
T_i=-\partial_{\theta_i} V
\end{equation}
where ${\bf r}_i$ and $\theta_i$ are the position and orientation of cell i. For conjugate gradient minimization, we transform these forces to a $3N$-dimensional gradient of the potential energy as $\nabla V=\{F_1^x,F_1^y,\rm M_1/I_1T_1, ..., F_N^x,F_N^y,\rm M_N/I_NT_N\}$, where the ratio of moment of inertia to mass of each cell is $\rm I_i/M_i=\frac{1}{8}\sigma^2\left(\frac{1+\Delta^4}{1+\Delta^2}+2\big(\frac{(1+\Delta)\Delta}{1+\Delta^2}\big)^2\right)$ with $\Delta_i=\frac{\sigma_{i,\rm bud}}{\sigma_{i,\rm mother}}$.

In this model, all mother cells have the same size, $\sigma_{i,{\rm mother}}=\sigma$. Cells grow in a square box with periodic boundary conditions. Cell growth progresses while $\sigma_{i,{\rm bud}}<\sigma$ and culminates in division. After division, both new cells acquire a random orientation (see Fig.~\ref{fig:cells}). 

\begin{figure*}
\includegraphics[width=0.95\textwidth,keepaspectratio]{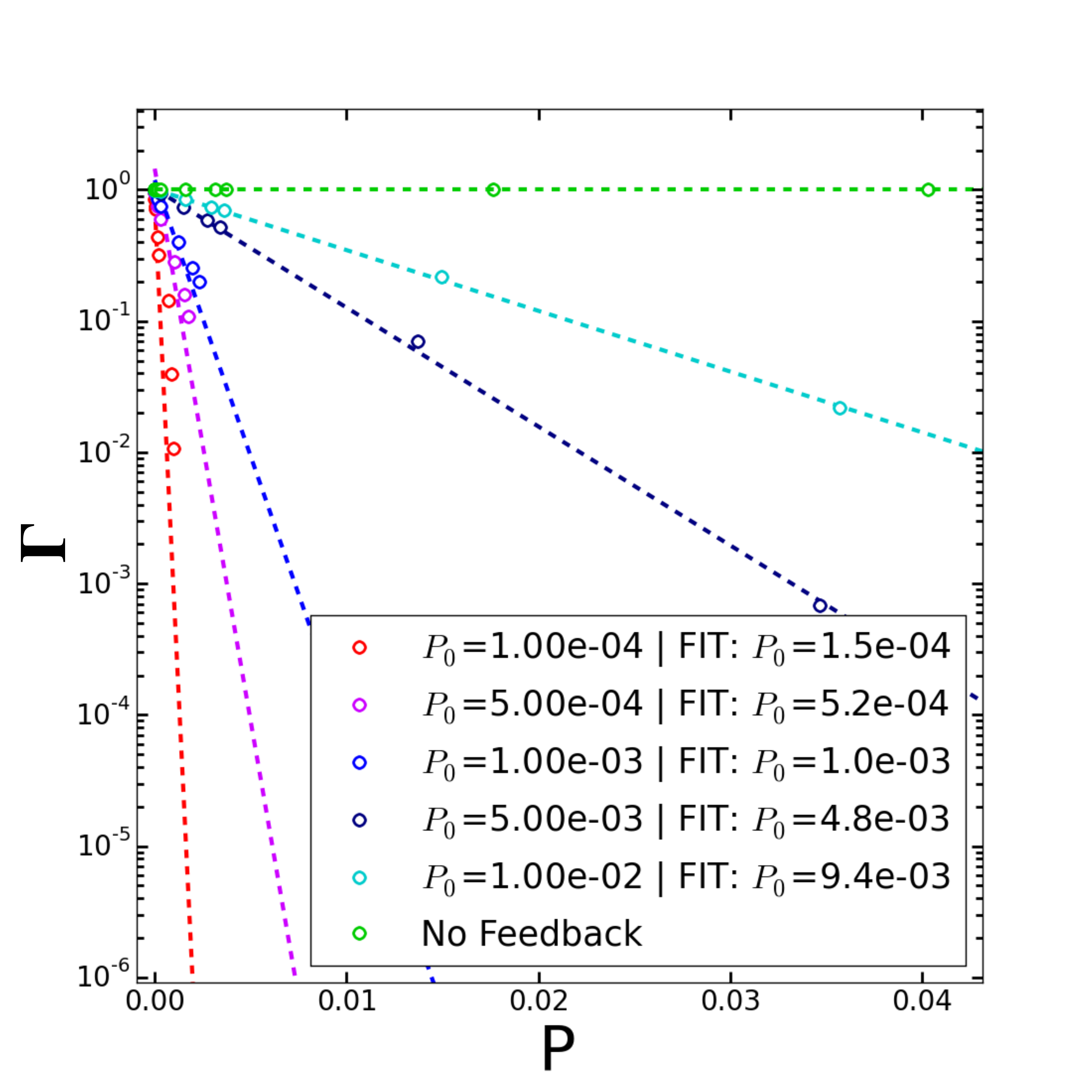}
\caption{Population-averaged growth rate $\Gamma$ as a function of the population averaged pressure $P$. $\Gamma$ is calculated from individual cell growth rates $\gamma_i$ as $\Gamma=1/A\sum_{i}\gamma(i)A(i)$, where $A(i)$ is the area of the $i^{th}$ cell, $\gamma(i)$ is the growth rate of the $i^{th}$ cell (at a given time-step), and $A = \sum_{i}A(i)$. Population-averaged pressure is calculated form the stress tensor $\Sigma_{\alpha\beta}$ (see Section \ref{P_and_G} for details). The results are for 6 different strengths of feedback: No feedback (green), $P_0=10^{-2}$(cyan), $P_0=5\cdot 10^{-3}$(dark blue), $P_0=10^{-3}$(blue), $P_0=5\cdot 10^{-4}$(purple), and $P_0=10^{-4}$ (red). Dashed-lines are fits to the numerical data. Fitted feedback strengths are given of the right-hand side of the legend. Simulations were done with time-steps kept constant, without the adaptive time-steps method. }
\label{fig:K_P}
\end{figure*}

Without pressure feedback, the growth rate for cell $i$ is: $\gamma_i=\gamma_i^0$ where $\gamma_i^0$ is chosen from a uniform distribution of width $20\%$ around a mean growth rate $\gamma^0$. With pressure feedback, the growth rate is additionally exponentially modulated as $\gamma_i=\gamma_i^0e^{-P_i/P_0}$ where $P_i$ is the pressure on the bud of the cell $i$, and $P_0$ is a strength of the feedback \cite{delarue2016self}. The pressure $\rm P_i$ is calculated as $\rm P_i = \sum_{j(i)}|F_{ij}|/L_{i,bud}$, where $\rm  L_{i,bud}=\pi\sigma_{i,bud}$ is the perimeter of bud $i$, $\rm |F_{ij}|$ is the magnitude of the contact force between a bud $i$ and a particle $j$, and $\rm j(i)$ is a set of the particles in contact with $i$. In Figure~\ref{fig:K_P} we show the relation between the strength of feedback $P_0$, population-averaged growth rate $\Gamma = \frac{1}{A}\sum_i a_i\gamma_i$ ($A=\sum_ia_i$), and growth-induced pressure $P$ (see Section \ref{P_and_G} for details on the calculation of $P$).

\subsection{Generating jammed packings}
The simulation starts with two randomly oriented yeast cells. Initially the cells grow according to Equation \ref{eq:growth} with a time-step of $\rm dt_0=0.002/\gamma^0$, and this continues while $\rm V/(kN) < \epsilon_{tol}^2$. If the population energy per cell at time step $\rm j$ is greater than $\rm V_j/(kN) > 2\cdot \epsilon_{tol}^2$, the growth step is rejected and the time step is halved ($\rm dt_{j}\rightarrow \frac{1}{2}\cdot\rm dt_{j}$), and the growth step (proceeded by energy minimization) is repeated (Equation \ref{eq:growth}). If the average potential drops below $\rm V/(kN) < \epsilon_{tol}^2$, the time step is reset to $\rm dt=\rm dt_0$. The simulation terminates when the the average energy per particles of a static energy is $\rm \epsilon_{tol}^2 < V/kN < 2\cdot \epsilon_{tol}^2$. 

This growth-driven protocol differs from previously-used compression protocols~\cite{olafsen2010experimental}. Whereas compression simulations start with a fixed number of objects and reduce the box size until the systems jams, growth-driven simulations start with few objects and increase the number of objects via growth until the system jams. 

Once the population reaches the jamming point at $\phi_J$, the colony grows beyond that point\textemdash up to the preassigned value $\delta \phi = \phi-\phi_J$. The protocol is similar to the one used to find $\phi_J$, however the time-step is halved when the volume fraction exceeds $\delta \phi$ by a margin of $\Delta=0.5\cdot 10^{-8}$, i.e. $\phi > \phi_J + \delta \phi + \Delta$. The time-step is reset to $dt=dt_0$ if the volume fraction falls below: $\phi < \phi_J + \delta \phi - \Delta$. The protocol ends when the volume fraction falls into the range $\phi \in \left[\phi_J + \delta\phi - \Delta , \phi_J+\delta\phi + \Delta \right]$.

To speed up simulations where population pressure significantly slows down growth ($P\gg P_0$), we use an adaptive time-step. In this method, we scale the time-step $dt$ by the largest cellular growth-rate in the population $\gamma_{\rm min}$: $dt'=dt\cdot \gamma/\gamma_{\rm min}$. This method ensures that, even while the population is under pressure, the fastest-growing cells add the same amount of volume per time-step as they would have without feedback. Unless noted explicitly, all results in this paper use this adaptive time-step method. 

\subsection{Calculation of mechanical properties}\label{P_and_G}

For each static packing, we calculate the stress tensor $\hat{\Sigma}$ via the Virial expression:
\begin{equation}
\rm \hat{\Sigma}_{\alpha\beta} = \frac{1}{2L^2}\sum_{i>j}\sum_{k,l}\big(r_{ik,jl}^\alpha F_{ik,jl}^\beta+r_{ik,jl}^\beta F_{ik,jl}^\alpha\big)
\end{equation}
where $\rm F_{ik,jlj}^\alpha$ is the $\alpha$-component of the force $\vec{F}_{ik,jl}$ on $k$th lobe of particle $i$ resulting from overlap with the $l$th lobe of particle $j$, and $\rm r_{ik,jl}^\alpha$ is the  $\alpha$-component of the vector from the center of mass of lobe $ik$ to the center of mass of lobe $jl$. The pressure is calculated from the stress tensor as $\rm P=\frac{1}{2}\big(\rm \hat{\Sigma}_{xx}+\rm \hat{\Sigma}_{yy}\big)$.

To calculate bulk modulus B, the simulation box is compressed by $d\phi=10^{-8}$, and the modulus is calculated from the definition $B=\phi dP/d\phi$.

To determine shear modulus G of a cellular packing, the response to quasistatic simple shear is calculated. To that end, for a static packing at $\delta \phi = \phi - \phi_j$, each cell is subject to a small affine shear strain (along the x direction with gradient in the y direction):
\begin{equation}
\rm x_i \rightarrow x_i + \delta\gamma y_i
\end{equation}
where $\rm \mathbf{r}_i = (x_i,y_i)$  is the location of the center of mass of a particle $i$, and $\delta \gamma = \delta x/L = 10^{-6}$, and $L$ is size of the system. Following the application of shear strain, the system is relaxed via energy minimization. Then, shear modulus is calculated from the definition $G=d\Sigma_{xy}/d\gamma$.

Throughout this paper we measure $P$, $B$, and $G$ in units of the cell-cell modulus $\rm k$.

\subsection{Pressure scale ($P_{\rm max}$) at confluency ($\phi=1$)}
\label{appx:pmax}

Population pressure $\rm P$ is determined by the overlap between cells, which in turn is set the compression beyond the jamming point $\delta\phi=\phi-\phi_J$. Here, we estimate the pressure $P_{\rm max}$ resulting from compression of cell packings from jamming $\phi_J=0.84$ to confluency $\phi=1$. 

By explicitly calculating the force between lobes $ik$ and $jl$, $F_{ik,jl}^\alpha=-\partial_{r_{ik}^\alpha}V=r_{ik,jl}^{\alpha}/r_{ik,jl}\times k\delta_{ik,jl}$, we can reduce the population pressure to $P=\frac{1}{2L^2}\sum_{i>j}\sum_{kl}kr_{ik,jl}\delta_{ik,jl}$. This can be expressed as an average over the $N_c=Nz/2$ contacts in the system, so that $P=\frac{Nz}{4L^2}\langle k r_{ik,jl}\delta_{ik,jl}\rangle$.

To estimate $P$, we assume that we have a population of monodisperse spherical cells with diameter $\sigma$ with $z\approx z_J=4$ and $\phi_J=0.84$. Further assuming that cells compress purely affinely upon infinitesimal compression, equivalent to swelling cells from $\sigma$ to $\sigma+d\sigma$ while keeping the box size fixed at $L$, and that no new contacts are made in the process allows us to simplify pressure to $P=\frac{N}{L^2}k\sigma d\sigma$. Since the volume fraction is $\phi=\frac{N\pi\sigma^2}{4L^2}$, $d\phi=\frac{N\pi\sigma}{2L^2}d\sigma$ and $\frac{N}{L^2}\sigma d\sigma=\frac{2}{\pi}d\phi$. This allows us to relate $P$ to changes in $\phi$: $P/k=\frac{2}{\pi}d\phi$.

Finally, we are able to calculated the pressure $P_{\rm max}/k=\frac{2}{\pi}0.16\approx 0.1$ generated by compressing a cell paacking from $\phi_J=0.84$ to $\phi=1$. This value is consistent with our measured data in Fig.~\ref{fig:sims}g.

\subsection{Modified isostaticity at jamming point}\label{appx:z-iso}

In this section we describe why the ``isostatic'' criterion for growing budding cells differs from the naive expectation of $Z_{\rm iso}^{\rm naive}=6$. 

\begin{figure*}
\includegraphics[width=\textwidth,keepaspectratio]{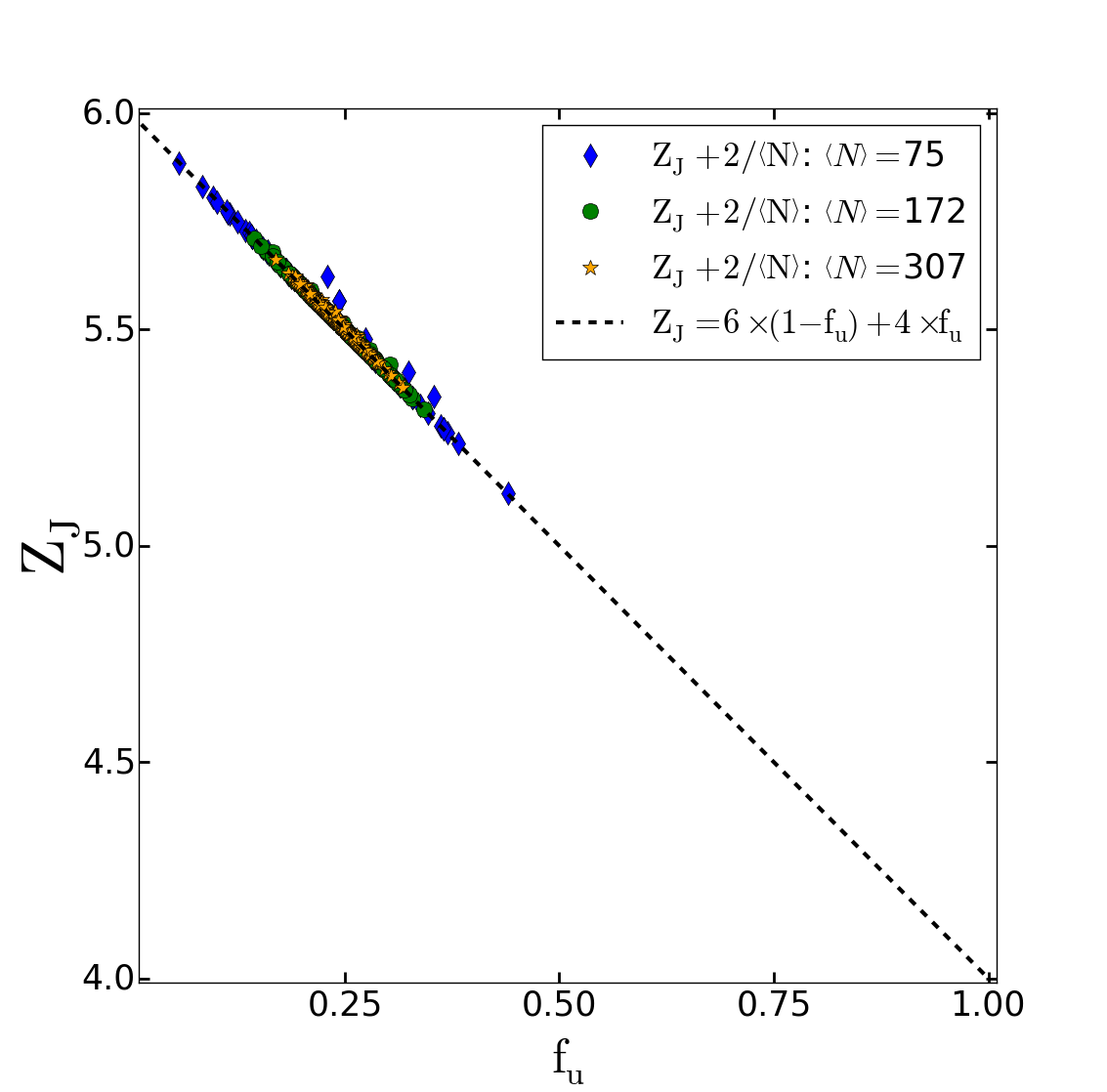}
\caption{
Average number of contacts $\rm Z_J$ as a function of a fraction of unconstrained buds $\rm f_{\rm u}$. Dashed-line gives the relation $Z_{\rm iso}=6-2f_u$ from constraint counting arguments. A small correction $2/\left<N\right>$ has been added to $\rm Z_J$ to account finite-size effects (Section \ref{appx:z-iso}).
All simulated packings have at least as many contacts as expected ($\rm Z_J\geq Z_{\rm iso}$) while the majority of packings exactly satisfy $\rm Z_J=Z_{\rm iso}$. Numerical data is shown for system sizes $\left<N\right>=75$,$\left<N\right>=172$ $\left<N\right>=307$.}
\label{fig:Z_iso}
\end{figure*}

In order for a system to be mechanically-stable, it must have as many contacts in the system ($N_c$) as degrees of freedom ($N_d$). This is the naive isostatic, or Maxwell, criterion. Since budding cells have $3$ degrees of freedom per cell, the system has a total of $N_d=3N$ degrees of freedom and we would naively expect there to be be $N_c=N_d=3N$ degrees of freedom at jamming, or $Z_{\rm iso}^{\rm naive}=2N_c/N=6$ contacts per cell. This argument breaks down, however, because not all degrees of freedom in the system are constrained. Unconstrained buds and rattlers decrease the contact number at jamming $Z_J$ below $Z_{\rm iso}^{\rm naive}=6$ because contacts are not required to constrain these degrees of freedom. The isostatic criterion, taking into account unconstrained buds and floaters, is $N_c^{\rm iso}=3N-3N_r-N_u-1$ where $N_r$ and $N_u$ are the number of rattlers and unconstrained buds in the system and the $-1$ is a finite-size correction.  We can express the isostatic criterion as $Z_{\rm iso}=2N_c^{\rm iso}/(N-N_r)=6-2f_u-2/(N-N_r)$ contact per (non-rattler) cell, where $f_u$ is the fraction of (non-rattler) cells with an unconstrained bud. Fig.~\ref{fig:Z_iso} shows that this isostacity criterion holds for nearly all packings analyzed. Note that in the main text we use the large-system limit $Z_{\rm iso}\big|_{N\rightarrow\infty}=6-2f_u$ and in Fig.~\ref{fig:Z_iso} we take into account finite-size effect by adding $2/(N-N_r)$ to $Z_J$ \cite{goodrich2012}. 

\subsection{Hyperstaticity due to unconstrained bud growth}
\label{appx:z-hyper}

In this section we describe why growth under extreme feedback produces ``hyperstatic'' ($Z>Z_{\rm iso}$) packings. In the case of extreme feedback ($P_0\rightarrow 0$), only unconstrained buds grow above the jamming point and these buds cease growing once they come into contact with their neighbors. If there are $N_u=f_uN$ unconstrained buds at jamming and of these $\Delta N_u=\Delta f_uN$ have come into contact with their neighbors due to growth above jamming, then the population needs $Z_{\rm iso}=2(3N-N_u+\Delta N_u)/N=6-2f_u+2\Delta f_u$ contacts for mechanical stability. However, each new unconstrained bud needs $2$ contacts to stabilize it, so that the contact number increases to $Z=2(3N-N_u+2\Delta N_u)/N=6-2f_u+4\Delta f_u$ as unconstrained buds make contact. So, as unconstrained buds make contact the system become hyperstatic with $\Delta Z'=Z-Z_{\rm iso}=2\Delta f_u$. In the extreme case, $\Delta Z'=2f_u$ is determined by the number of buds that were unconstrained at jamming. Note that this deviation from isostaicity is twice as large as the deviation from the coordination number at jamming, $\Delta Z=Z-Z_J=4\Delta f_u$, which we show in the paper.  

\begin{figure*}
\includegraphics[width=\textwidth,keepaspectratio]{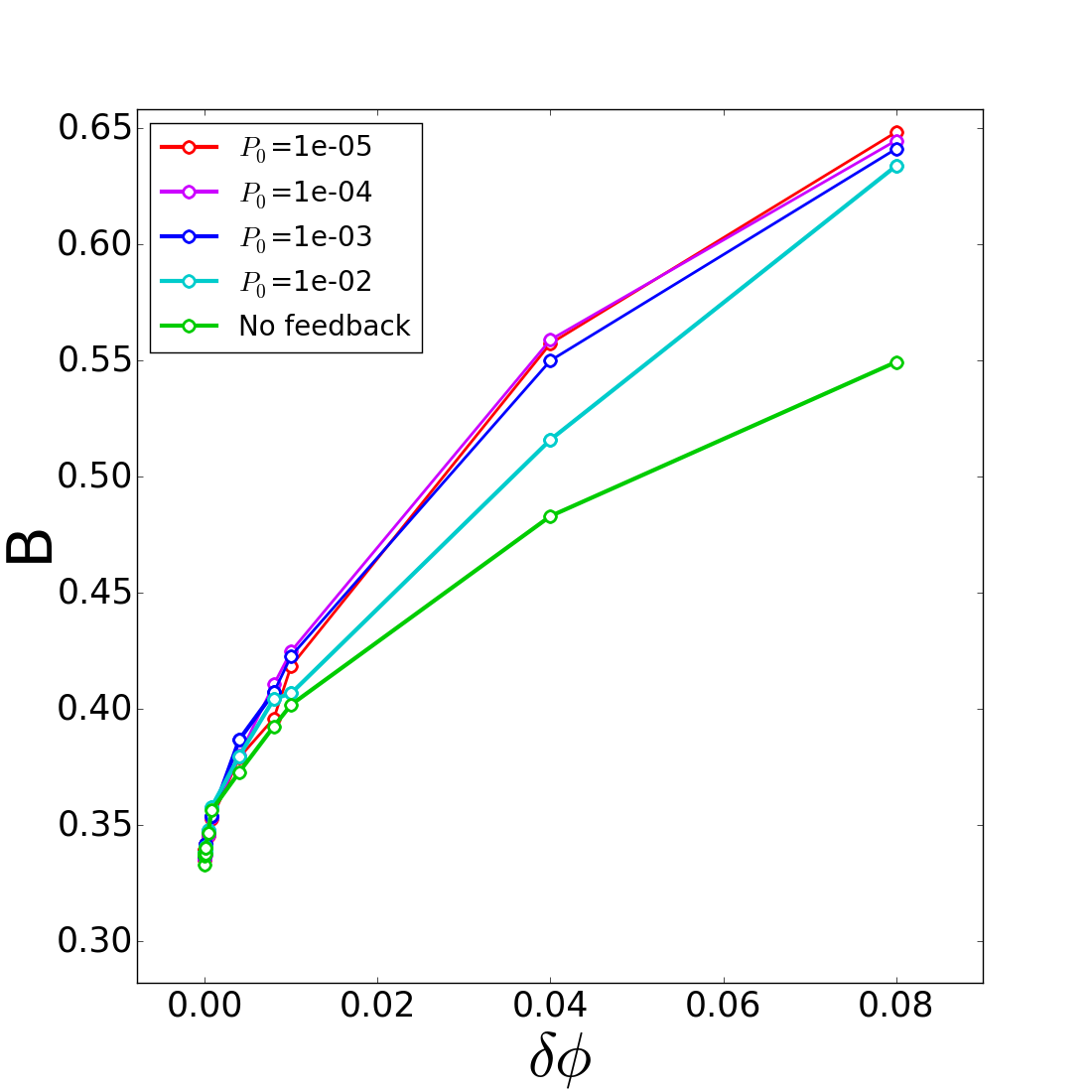}
\caption{
Bulk modulus as a function of volume fraction above jamming $\delta\phi=\phi-\phi_J$ for populations growing under five different feedback strengths. 
}
\label{fig:B_dPhi_linear}
\end{figure*}

\begin{figure*}
\includegraphics[width=\textwidth,keepaspectratio]{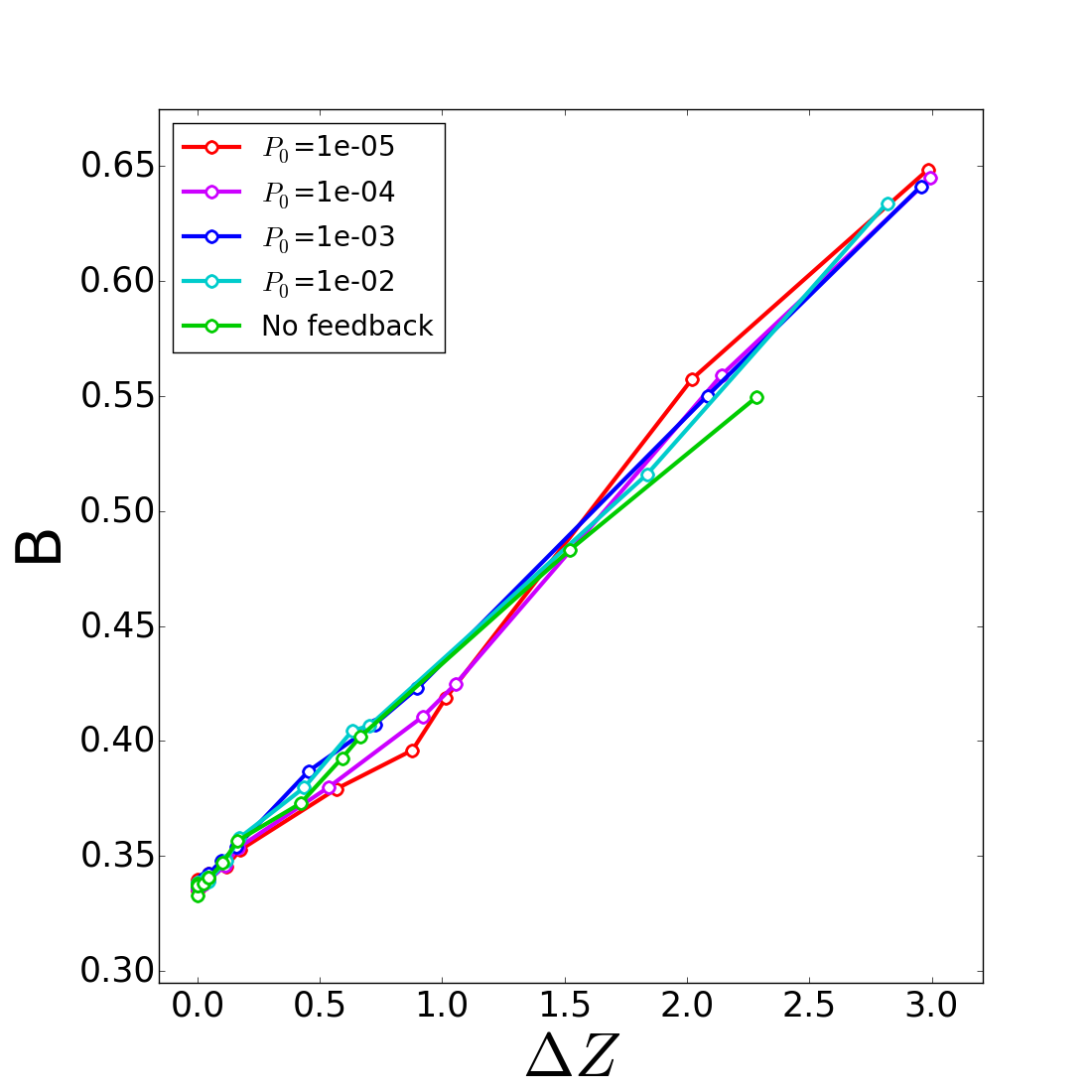}
\caption{
Bulk modulus as a function of number of  cell-cell contacts above jamming $\Delta Z=Z-Z_J$ for populations growing under five different feedback strengths. 
}
\label{fig:B_dZ_linear}
\end{figure*}

\end{document}